\documentclass[runningheads]{llncs}
\usepackage{graphicx}
\usepackage{todonotes}
\presetkeys{todonotes}{inline}{}
\begin{document}
\title{Open Data Chatbot}

%
%
\author{Sophia Keyner \and Vadim Savenkov \and Svitlana Vakulenko}
\institute{Vienna University of Economics and Business \\ Institute for Information Business \\ \email{sophia.keyner@s.wu.ac.at} \\
\email{vadim.savenkov@wu.ac.at} \\
\email{svitlana.vakulenko@wu.ac.at}}

%
\maketitle              
\begin{abstract}
Recently, chatbots received an increased attention from industry and diverse research communities as a dialogue-based interface providing advanced human-computer interactions.
On the other hand, Open Data continues to be an important trend and a potential enabler for government transparency and citizen participation.
This paper shows how these two paradigms can be combined to help non-expert users find and discover open government datasets through dialogue.

\keywords{Open Data  \and Conversational search.}
\end{abstract}
\section{Introduction}


Open Data is often used to create new services and applications. 
The major source of Open Data, remain government Open Data portals designed to provide more transparency and enable the general public to monitor the state of affairs and exercise control over the actions of government bodies.
Open Data access is expected to encourage public awareness and citizen participation.


Chatbot is a software providing a conversational interface.
One of the applications for chatbots in conversational search providing access to an information source, such as a database.
Chatbots have a number of advantages as well as associated challenges in comparison with other graphical interfaces:
\begin{itemize}
\item Light-weight and inexpensive to deploy, especially in comparison with custom visualizations;
\item Engaging and intuitive, do not require user training.
\end{itemize}

In this paper we present an implementation of a chatbot that provides conversational search interface to an Open Data repository utilizing geo-entity annotations.
Our code is open-source\footnote{\texttt{https://git.ai.wu.ac.at/sophiakeyner/open\_data\_chatbot}} and the web interface is publicly available\footnote{\texttt{https://bot-test.communidata.at}}.

\section{Related Work}

Chatbots were introduced as an alternative interface able to provide a more user-friendly and engaging experience, while also serving as an effective information access point to structured data sources, such as Open Data repositories~\cite{DBLP:conf/esws/HeilN18,DBLP:conf/esws/NeumaierSV17,DBLP:conf/icwe/PorrecaLMVC17} and knowledge graphs~\cite{DBLP:conf/www/AthreyaNU18}.
DBpedia chatbot~\cite{DBLP:conf/www/AthreyaNU18}, for example, relies on a handful of pre-defined rules, which prevents from a more fine-grained query analysis, requires continuous engineering efforts and makes the approach difficult to scale and incorporate new conversational patterns.
The approach we describe in this paper leverages machine-learning models that can be further improved by continuously learning from interactions with end-users.
We show how machine-learning-based dialogue systems can benefit from explicit semantics, such as geographic entities (locations) annotations, that assist in natural language interpretation.

Open Data makes a perfect use case for the integration of the spatial linked data source, as it was previously shown to heavily rely on the geographic dimensions to enable effective search and discovery~\cite{DBLP:conf/www/KacprzakKTS18,DBLP:conf/i-semantics/NeumaierSP18}.
We leverage these insights and demonstrate how semantic geo-information linked data sources can help to enhance user experience with automated dialogue systems.
Our work is a follow-up on the previously proposed Open Data chatbot architectures~\cite{DBLP:conf/esws/HeilN18,DBLP:conf/esws/NeumaierSV17}.
The main difference to the previously proposed architecture~\cite{DBLP:conf/esws/NeumaierSV17} is providing support for dataset discovery functionality beyond conversational search towards enabling conversational browsing of the underlying data structures.
The mode switch is triggered by the intent detection component of our system.

Our implementation also utilizes the results of the semantic annotation approach proposed by Neumaier et al. \cite{DBLP:conf/i-semantics/NeumaierSP18} similar to the chatbot for geo-search and visualization by Heil and Neumaier \cite{DBLP:conf/esws/HeilN18}.
We make a step further by training a supervised model for intent detection and entity recognition from the user natural-language input.

\section{Chatbot Architecture}

\begin{figure}[ht]
\centering
\includegraphics[width=8.5cm]{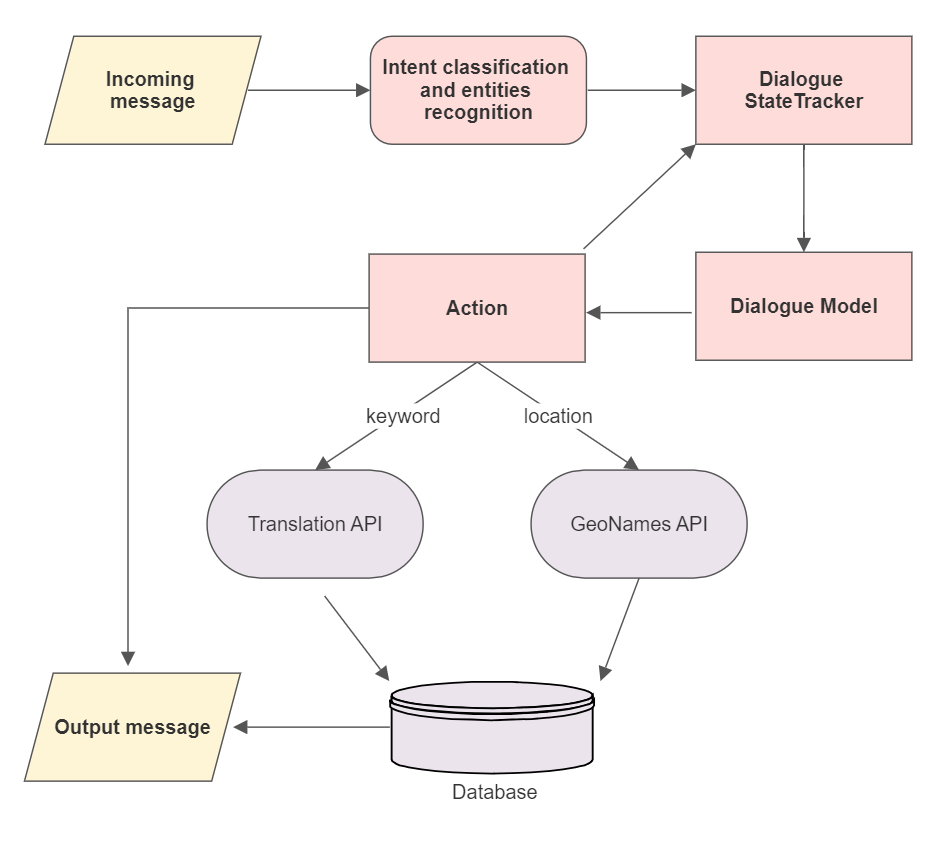}
\caption{Open Data Chatbot architecture.}
\label{fig:rasa}
\end{figure}

Our architecture design is based on the Rasa framework\footnote{\texttt{https://rasa.com}}, which also provides an open-source Python library that implements several models for training customized dialogue systems.
The chatbot architecture integrates different components into a single processing pipeline that takes an input from the user and produces a response.
Figure~\ref{fig:rasa} depicts the main steps of the processing pipeline across the components.



\subsection{Message Interpretation}

The message interpretation module includes separate machine learning models for entity recognition and intent classification tasks. 
We hand-crafted 250 sample messages to train these models.

\textbf{Entity Recognition}.
Our proof-of-concept prototype can recognize two types of entities: topic keywords and geo-entities.
The first step, entity mention extraction, is an instance of the sequence labeling task~\cite{DBLP:conf/icml/LaffertyMP01}.
Its input is the text of the user utterance represented as a sequence of words.
Then, a supervised machine learning model, conditional random fields (CRF) in our implementation, is trained to assign labels to words.
121 of the sample messages, which were used for training the model, included one of the topic keywords and 18 samples contained one of the geo-entities.
These labels are then used to extract entity mentions.
By entity mention we mean a text span that refers to one of the entities, e.g. ``schools'' or ``Graz''.
Entity mentions are then used in the search query to retrieve and rank relevant datasets from our Open Data repository.
We found out that the pre-trained model often fails to extract geo-entities that were absent from the training dataset.
To mitigate this issue we implemented a look-up table that contains a list of geo-entities as an alternative unsupervised approach for entity mention extraction.



\textbf{Intent Classification} was trained with a support vector machine (SVM) classifier to recognize nine intents: greeting, good-bye, add keyword, add location, search, explore, thank you, affirm, deny.
We designed at least six sample messages for each of the intents.

\subsection{Dialogue Management}
The sequence of selected actions is tracked and determines the current dialogue state.
Our set of available actions includes 10 hand-crafted utterance-templates and a custom action to access the database.
The dialogue management component receives the entities and the intent detected at the previous step of message interpretations and selects the next action from the predefined set of actions.
This selection is made by a neural network model, which was trained on 14 hand-crafted stories, based on the user's intent and the current dialogue state.




\section{Dialogue Flow}

\begin{figure}[ht]
\centering
\includegraphics[width=13cm]{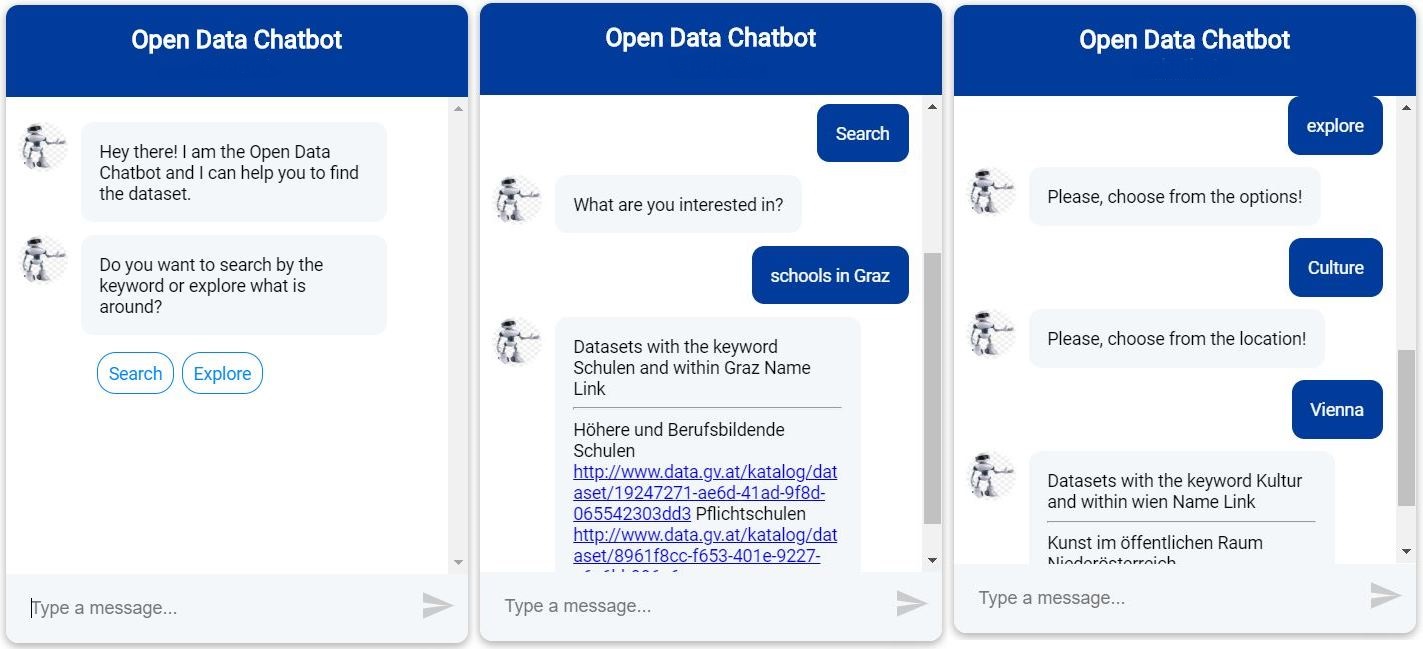}
\caption{Example of the search and exploratory modes.}
\label{fig:screens}
\end{figure}

After greeting the user, Chatbot requests User to make a choice between two interaction modes: search and explore (see Figure~\ref{fig:screens}).
The mode can be switched at any time with an ad-hoc user utterance (e.g. "Could I go back to explore?") that should be classified with the corresponding intent.

In search mode, Chatbot requests User to define the topic of interest.
In exploratory mode, Chatbot asks two follow-up questions which the available options displayed as buttons.
The first question is the topic of interest, such as education and health care, the second is the geo-location that defines the scope of the dataset.
User then can choose one of the options provided by Chatbot or enter a custom query instead.
Then, Chatbot answers with at most five matching dataset titles and links to the original Open Data portal.




\section{Conclusion}
We implemented a novel prototype of an Open Data chatbot by integrating state-of-the-art parsing and semantic technologies.
This demonstration showcases a conversational search application for a repository of publicly available datasets, but the design approach we described is not limited to this specific use case and can be transferred to other domains, such as cultural heritage or e-commerce, where users can benefit from more engaging and compact interfaces for search and discovery.
The main limitation for training dialogue models using machine learning techniques is the need for annotated conversational data.
We showed that it is possible to develop such models using only a handful of examples, but it is important to look further for more scalable solutions that can help to gradually extend these models beyond the initial built-in assumptions about the dialogue structure.

\paragraph{\bf Demonstration plan.}
We will engage conference participants in the evaluation of our system designed as a user study.
The task is given a broad topic, such as Nature, find relevant datasets using our chatbot interface.
The participants will be encouraged to fill out a small questionnaire in the end, which includes rating of their experience, providing additional feedback and a wish list.

\noindent
\textbf{Acknowledgements.}
This work was supported by the Austrian Research Promotion Agency (FFG) under the project CommuniData (grant no. 855407).

%
%
%
\bibliographystyle{splncs04}
\bibliography{mybibliography}
\end{document}